\pdfoutput=1
\documentclass[amsmath,amssymb,aps,,pra,reprint,superscriptaddress]{revtex4-2} 

\usepackage[svgnames,psnames]{xcolor}
\usepackage{tabularx}
\usepackage{url}
\usepackage{braket}
\usepackage{graphicx}% Include figure files
\graphicspath{{.}{figs/}{../figs/}}
\usepackage{afterpage}
\usepackage{dcolumn}% Align table columns on decimal point
\usepackage{bm}% bold math
\usepackage{tikz}
\usetikzlibrary{positioning,intersections,angles,quotes}
\usepackage{comment}
\usepackage{physics}
\usepackage{mathtools}
\usepackage[colorlinks,citecolor=DarkGreen,linkcolor=FireBrick,linktocpage,unicode,hypertexnames=false]{hyperref}
\usepackage{ulem}
\usepackage{mathcomp}
\usepackage{quantikz}
\newcommand{\T}{\hat{T}}
\newcommand{\Td}{\hat{T}^\dagger}

\usepackage{enumerate}

\usepackage{subcaption}
\usepackage{graphicx}
\usepackage{enumerate}
\begin{document}

\title{Chebyshev Approximated Variational Coupled Cluster for Quantum Computing}
\author{Luca Erhart}
\email{erhartl@hotmail.com}
\affiliation{%
  Graduate School of Engineering Science, Osaka University, 1-3 Machikaneyama, Toyonaka, Osaka 560-8531, Japan
}%
\author{Yuichiro Yoshida}%
\email{yoshida.yuichiro.qiqb@osaka-u.ac.jp}
\affiliation{%
  Center for Quantum Information and Quantum Biology,
  Osaka University, 1-2 Machikaneyama, Toyonaka 560-8531, Japan
}%

\author{Viktor Khinevich}
\affiliation{%
  Graduate School of Engineering Science, Osaka University, 1-3 Machikaneyama, Toyonaka, Osaka 560-8531, Japan
}%

\author{Wataru Mizukami}%
\email{mizukami.wataru.qiqb@osaka-u.ac.jp}
\affiliation{%
  Graduate School of Engineering Science, Osaka University, 1-3 Machikaneyama, Toyonaka, Osaka 560-8531, Japan
}%
\affiliation{%
  Center for Quantum Information and Quantum Biology,
  Osaka University, 1-2 Machikaneyama, Toyonaka 560-8531, Japan
}%

\date{\today}
	
\begin{abstract}
We propose an approach to approximately implement the variational coupled cluster (VCC) theory on quantum computers, which struggles with exponential scaling of computational costs on classical computers. To this end, we employ expanding the exponential cluster operator using Chebyshev polynomials and introduce two methods: the Chebyshev approximated VCC (C$^d$-VCC) and the Hermitian-part Chebyshev approximated VCC (HC$^d$-VCC), where $d$ indicates the maximum degree of the Chebyshev polynomials.  The latter method decomposes the cluster operator into anti-Hermitian and Hermitian parts, with the anti-Hermitian part represented by the disentangled unitary coupled cluster ansatz and the Hermitian part approximated using Chebyshev expansion. We illustrate the implementation of the HC$^d$-VCC in a quantum circuit using the quantum singular value transformation technique. Numerical simulations show that the C$^d$-VCC rapidly converges to the exact VCC with increasing truncation degree $d$, and the HC$^d$-VCC effectively reduces the Chebyshev expansion error compared to the C$^d$-VCC. 
The HC$^d$-VCC method for realizing non-unitary coupled cluster wave functions on quantum computers is expected to be useful for initial state preparation on quantum computers and efficient tomography of the quantum state for post-processing on classical computers after quantum computations.
\end{abstract}

\maketitle

\section{Introduction}
Coupled cluster (CC) theory is a highly effective framework for quantum chemical calculations~\cite{Coupled_cluster}. For weakly correlated systems, the coupled cluster with single, double, and perturbative triple excitations (CCSD(T)) is widely known to achieve chemical accuracy. The development of CC methods has progressed steadily within the quantum chemistry community, making them applicable to a wide range of systems. In 2013, a CC calculation of an entire small protein was achieved~\cite{Neese2014}. In recent years, applications have extended beyond molecular systems to include solid-state and surface systems under periodic boundary conditions~\cite{PhysRevX.8.021043, Zhang_2019}. Coupled cluster with single and double excitations (CCSD) and CCSD(T) are routinely used as reference methods, and have been actively utilized for creating datasets for machine learning, especially in recent times~\cite{JSmith2019NatCommun,DWilkins2019PNAS,KSorathia2024JCTC}.

However, it is well known that CC methods do not work well for strongly correlated systems. CC methods assume weak electron correlations and are not variational; therefore, they break down for complicated electronic structures. If CC methods could be solved variationally, their range of applicability would greatly expand. Solving variational CC (VCC)~\cite{Bartlett1988,Kutzelnigg1991,Szalay1995,Cooper2010,Knowles2010,Evangelista2011,Robinson2011,Robinson2012,Harsha2017,VanVoorhis2000,Marie2021} on a classical computer requires an exponential computational cost, making VCC calculations virtually impossible for most molecules.

This situation changed in 2014 when it was revealed that unitary coupled cluster (UCC) theory~\cite{Unitary_coupled_cluster}, a type of VCC, could be solved in polynomial time using a quantum-classical hybrid algorithm now known as the variational quantum eigensolver (VQE)~\cite{Peruzzo2014}. This study triggered extensive research on UCC and related theories using quantum computers. Over the past decade, numerous studies has revealed several practical problems with VQE~\cite{VQE_review}. One major issue is the enormous measurement cost (sampling cost) required to calculate energies with the accuracy needed for chemistry~\cite{PhysRevResearch.4.033154}. Another problem is the lack of efficient algorithms for optimizing the parameters of nonlinear wavefunction models. Parameter optimization requires repeated energy calculations, and a large number of optimization steps means that high-cost energy calculations must be repeated, making parameterization of wavefunctions such as UCC using VQE difficult~\cite{AGu2021arXiv}.

To address this problem, it is desirable to determine as many parameters as possible in advance using a classical computer. This requires a wavefunction ansatz that can be optimized on a classical computer. For example, Matrix Product States~\cite{RevModPhys.93.045003} can be optimized classically and are easy to implement on quantum computers. However, UCC is different from classical CC, and while some approximate parameters can be prepared on a classical computer, it is difficult to ensure sufficient correspondence. If CC could be directly implemented on a quantum computer, this issue would be greatly improved.

Another issue is that VQE and quantum phase estimation (QPE)~\cite{Abrams1999PRL,Aspuru2005Science} generally require more than twice the number of qubits as spatial orbitals. When using basis functions of the size needed for quantitative calculations, the number of required qubits becomes very large. Therefore, in VQE and QPE, it is inevitable that the application of quantum computers will be limited to essential degrees of freedom, necessitating the introduction of the active space orbital approximation. The electron correlation in other degrees of freedom outside the active space must be considered using a classical computer before or after the quantum computation. In this case, the information from the wavefunction on the quantum computer needs to be transferred to the classical computer. However, measuring high-order reduced density matrices (RDMs) is unrealistic due to the enormous measurement cost~\cite{Gonthier2022PRR,Tilly2021PRR,Nishio2023PCCP, takemori2023balancing}, and approximate tomography methods~\cite{Huang2020, Kohda2022} also require repeating a considerable number of measurements. In contrast, if a common ansatz could be used for both classical and quantum computers, such as implementing CC on a quantum computer, the wavefunction information could be directly transferred to the classical computer from the parameters of the quantum circuit. Therefore, realizing the widely used non-unitary CC on classical computers as a quantum circuit, which connects quantum and classical methods, is significant.

In this study, we aimed to implement non-unitary CC theory on quantum computers. Specifically, we proposed an approximated VCC theory by expanding the exponential ansatz in Chebyshev polynomials. We also developed a method for Chebyshev expansion based on the decomposition of the cluster operator into Hermitian and anti-Hermitian parts, and showed its implementation in a quantum circuit. 

The structure of this paper is organized as follows: In Section II, we provide an overview of the VCC theory and the Chebyshev expansion method. We explain the method of Chebyshev expansion based on Hermitian cluster operators and its implementation in a quantum circuit. In Section III, we detail the numerical simulations, including the software, algorithms, and computational conditions used. In Section IV, we present the results of the numerical verification for small molecules and evaluate the accuracy of VCC using Chebyshev expansion. Finally, in Section V, we summarize this study and discuss future prospects.

\section{Methods} \label{seq_theory}

\subsection{Review of variational coupled cluster} \label{sec_VCC}

VCC~\cite{Bartlett1988,Kutzelnigg1991,Szalay1995,Cooper2010,Knowles2010,Evangelista2011,Robinson2011,Robinson2012,Harsha2017,VanVoorhis2000,Marie2021} is a variant of CC. The energy expression of VCC is given by
\begin{align} \label{eq_E_vCC}
 E_{{\rm VCC}} := \frac{\bra{\Psi_0} e^{\hat{T}^\dagger} \hat{H} e^{\hat{T}} \ket{\Psi_0}}{\bra{\Psi_0} e^{\hat{T}^\dagger}e^{\hat{T}} \ket{\Psi_0}},
\end{align}
where $\ket{\Psi_0}$ is the reference wave function, usually the Hartree-Fock (HF) state. The operators $\hat{H}$ and $\hat{T}$ are the electronic Hamiltonian and the cluster operator. The cluster operator $\hat{T}$ comprising single and double excitation operators is expressed as
\begin{align} \label{eq_T_ccsd}
    \hat{T} = \sum_{i,a} t_i^a \hat{a}_a^\dagger \hat{a}_i 
    + \sum_{i<j,\,a<b} t_{ij}^{ab} \hat{a}_a^\dagger \hat{a}_b^\dagger \hat{a}_i \hat{a}_j,
\end{align}
where $\hat{a}^\dagger_a$ and $\hat{a}_i$ are the creation and annihilation operators of the $a$-th and $i$-th orbitals, respectively. 
The indices $i$, $j$, $\cdots$ and $a$, $b$, $\cdots$ correspond to the occupied and unoccupied orbitals, respectively.
The coefficients $t_i^a$ and $t_{ij}^{ab}$ are known as the cluster amplitudes. The exponential ansatz with the single and double cluster operators is called the coupled cluster singles and doubles (CCSD) ansatz. 

In variational coupled cluster singles and doubles (VCCSD), these parameters are variationally optimized to minimize the energy of Eq.~(\ref{eq_E_vCC}). However, computing the VCC energy on classical computers is known to require exponential computational costs. This is because the Baker–Campbell–Hausdorff expansion of $\bra{\Psi_0} e^{\hat{T}^\dagger} \hat{H} e^{\hat{T}} \ket{\Psi_0}$ is not terminated. To address this issue, the standard CC method employs the following energy expression based on the similarity transformation:
\begin{align} \label{eq_E_CC}
 E_{{\rm CC}} := \bra{\Psi_0} e^{-\hat{T}} \hat{H} e^{\hat{T}} \ket{\Psi_0}.
\end{align}
This approach allows the computation of CC energy and the determination of the cluster amplitudes with polynomial costs on classical computers.
One drawback of this method is that the energy is no longer variational and breaks down when the overlap between the reference wavefunction and the target wavefunction is small.

\subsection{Chebyshev expansion of VCC towards quantum computing applications}

Evaluating $E_{\rm VCC}$ is challenging, even using quantum computers, due to the non-unitary nature of $e^{T}$. Here, we introduce the Chebyshev expansion of $e^{T}$ for implementation on a quantum computer, where QSVT is employed to prepare Chebyshev polynomials.

\subsubsection{Chebyshev expansion of $e^{\hat{T}}$}
The exponential ansatz $e^{\hat{T}}$ can be expanded using Chebyshev polynomials:
\begin{equation} \label{eq_chev_expand}
e^{\T} = \sum_{n=0}^{N} c_n \mathcal{T}_n(\T/\tau) , 
\end{equation}
where $\mathcal{T}_n$ and $c_n$ are the $n$-th degree Chebyshev polynomials of the first kind and their real coefficients, respectively.
$N$ denotes the number of electrons. 
Since the Chebyshev expansion of a function $f(x)$ is only defined for $|x| \le 1$, the cluster operator $\hat{T}$ is normalize with $ \tau = |\T|$.
The Chebyshev polynomial coefficients $c_n$ are determined for the Chebyshev expansion of $f(x) = (e^\tau)^x$ to compensate for the normalization of the cluster operator $\hat{T}$.

We truncate the Chebyshev polynomial expansion in Eq.~(\ref{eq_chev_expand}) at $n=d$ to reduce computational costs. Henceforth, we will denote the Chebyshev polynomial expansion of $e^{\hat{T}}$ truncated at the $d$-th degree as $\hat{A}^{(d)}(\hat{T})$.
For example, the low-degree truncated ansatzes of the Chebyshev polynomial expansion of $e^{\hat{T}}$ are represented as
\begin{align}
    \hat{A}^{(0)}(\hat{T}) &= c_0, \\
    \hat{A}^{(1)}(\hat{T}) &= c_0 + c_1 \hat{T}/\tau, \\
    \hat{A}^{(2)}(\hat{T}) &= (c_0 - c_2) + c_1 \hat{T}/\tau + 2 c_2 (\hat{T}/\tau)^2.
\end{align}
In this manuscript, we refer to the Chebyshev expansion of VCCSD with degree $d$ as C$^d$-VCCSD.
The ansatzes $\hat{A}^{(d)}(\hat{T})$ truncated at low degrees correspond to well-known quantum chemical theories.
By substituting the $0$-degree ansatz $\hat{A}^{(0)}(\hat{T})$ for $e^{\hat{T}}$ into Eq.~(\ref{eq_E_vCC}), it is evident that the energy expectation value of the HF state is obtained.
Applying the $1$-degree ansatz $\hat{A}^{(1)}(\hat{T})$ enables the wave function to incorporate single and double excitations and assume the same form as configuration interaction with singles and doubles (CISD):
\begin{align} \label{eq:Cheby1WF}
    &\hat{A}^{(1)}(\hat{T})\ket{\Psi_0} \notag \\
    %&= (c_0 + c_1 \hat{T})\ket{\Psi_0} \\
    %&= (c_0 + c_1 \sum_{i,a} t_i^a \hat{a}_a^\dagger \hat{a}_i 
    %+ c_1 \sum_{i<j,a<b} t_{ij}^{ab} \hat{a}_a^\dagger \hat{a}_b^\dagger \hat{a}_i \hat{a}_j)\ket{\Psi_0} \\
    &\quad= c_0 \ket{\Psi_0} + c_1/\tau \sum_{i,a} t_i^a \ket{\Psi_{i}^{a}} + c_1/\tau \sum_{i<j,a<b} t_{ij}^{ab} \ket{\Psi_{ij}^{ab}},
\end{align}
where $\ket{\Psi_{i}^{a}}$ and $\ket{\Psi_{ij}^{ab}}$ are the singly and doubly excited configurations, respectively.

\subsubsection{Chebyshev expansion of $e^{\hat{T}}$ based on a decomposed Hermitian cluster operator}

The implementation of the truncated Chebyshev expansion of the VCC ansatz on a quantum computer requires an additional approximation. 
QSVT does not allow for making a polynomial of any matrix on quantum computers. The matrix for QSVT should be Hermitian. However, the cluster opertor $\hat{T}$ is nilpotent, making QSVT not directly applicable for constructing $\mathcal{T}_n(\T/\tau)$.

To address this issue, we consider the following approximated form of the VCC ansatz:
\begin{equation} \label{eq_one_trotter_step}
	e^{\T} = e^{\frac{1}{2}(\T -\Td) + \frac{1}{2}(\T+\Td)} \approx e^{\frac{1}{2}{(\T - \Td)}}	e^{\frac{1}{2}{(\T + \Td)}}.
\end{equation}
In this approach, the cluster operator is decomposed into the anti-Hermitian $\frac{1}{2}(\T -\Td)$ and Hermitian parts $\frac{1}{2}(\T+\Td)$. In this study, we call the ansatz as a {\it trotterized} VCC ansatz. The exponential function of the anti-Hermitian part $\frac{1}{2}(\T -\Td)$ is unitary and closely resembles the UCC theory. The implementation of a UCC ansatz in a quantum circuit has been extensively researched. The disentangled UCC ansatz~\cite{Evangelista2019, Taube2006} can be utilized to implement the former part:
\begin{equation} \label{eq:ducc}
e^{\frac{1}{2}{(\T - \Td)}} \approx 
\prod_{\mu} e^{\frac{1}{2}(T_{\mu} - \Td_{\mu})},
\end{equation}
where $\hat{T}_\mu$ denotes each term of the cluster operator $\hat{T}$ such as $t_{i}^{a} \hat{a}_a^\dagger  \hat{a}_i$ and $t_{ij}^{ab} \hat{a}_a^\dagger \hat{a}_b^\dagger \hat{a}_i \hat{a}_j$.

The Chebyshev expansion is applied to the latter part $e^{\frac{1}{2}{(\T + \Td)}}$. As mentioned in the previous subsection, we normalize the operator $\frac{1}{2}{(\T + \Td)}$ with $\kappa=|\frac{1}{2}{(\T + \Td)}|$ to ensure the definition of the Chebyshev expansion. The resulting Chebyshev expansion truncated at $n=d$ is expressed as:
\begin{equation} \label{eq:Cheby_Hermitian_part}
e^{\frac{1}{2}{(\T + \Td)}} \approx \sum_{n=0}^{d} c_n \mathcal{T}_n\left(\frac{1}{2}{(\T + \Td)/\kappa}\right).
\end{equation}

An approximate VCC ansatz that can be implemented in a quantum circuit is as: 
\begin{equation} \label{eq:hc_vcc_ansatz}
e^{T}
\approx 
\prod_{\mu} e^{\frac{1}{2}(T_{\mu} - \Td_{\mu})}
\left(
\sum_{n=0}^{d} c_n \mathcal{T}_n\left(\frac{1}{2}{(\T + \Td)/\kappa}\right)
\right)
\end{equation}
We refer to this ansatz as the Hermitian-part Chebyshev approximation with degree $d$ of VCC or HC$^d$-VCC in the remainder of this paper.

\subsection{The QSVT implementation of HC$^d$-VCC ansatz}

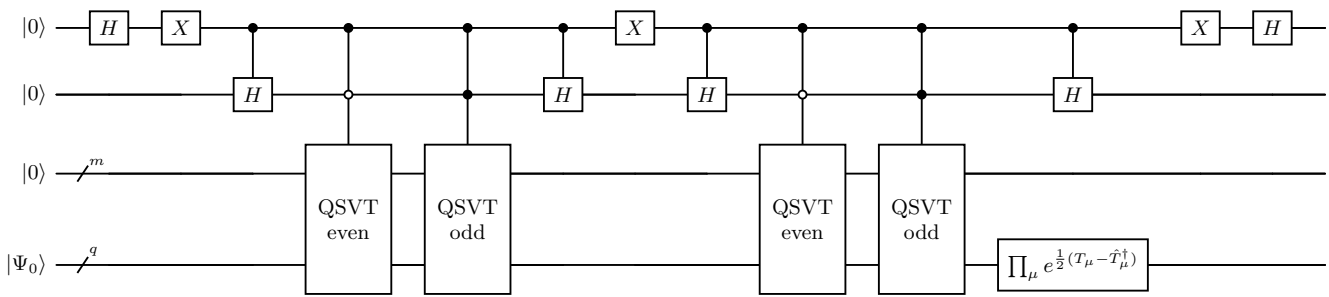
\begin{figure*}[tb]
\centering
\resizebox{\textwidth}{!}{
\begin{quantikz}
\lstick{$\ket{0}$} & \gate{H} & \gate{X} & \ctrl{1} & \ctrl{2} & \ctrl{2} & \ctrl{1} & \gate{X} & \ctrl{1} & \ctrl{2}& \ctrl{2} & \ctrl{1} & \gate{X}& \gate{H} & \qw \\
\lstick{$\ket{0}$}& \qw& \qw & \gate{H} & \ocontrol{} & \control{} & \gate{H}& \qw& \gate{H}& \ocontrol{}& \control{}& \gate{H} & \qw& \qw & \qw \\
\lstick{$\ket{0}$}&\qwbundle{m} & \qw & \qw & \gate[wires=2]{\begin{minipage}{1cm}\centering QSVT \\ even\end{minipage}} & \gate[wires=2]{\begin{minipage}{1cm}\centering QSVT \\ odd\end{minipage}} & \qw & \qw & \qw & \gate[wires=2]{\begin{minipage}{1cm}\centering QSVT \\ even\end{minipage}} & \gate[wires=2]{\begin{minipage}{1cm}\centering QSVT \\ odd\end{minipage}}& \qw & \qw& \qw& \qw  \\
\lstick{$\ket{\Psi_0}$}&\qwbundle{q} & \qw & \qw & & \qw & \qw & & \qw & \qw & & \gate{\prod_{\mu} e^{\frac{1}{2}(T_{\mu} - \Td_{\mu})}} & \qw & \qw & \qw 
\end{quantikz}
}
\caption{Quantum circuit to execute the HC$^d$-VCC ansatz for a $q$-qubit system. QSVT even and QSVT odd apply the corresponding part of the Chebychev polynomial with even/odd parity, respectively. The block encoding of $\frac{1}{2}(\T + \Td) $ in QSVT is performed using $m$-ancilla qubits. A LCU structure with two ancilla qubits is required to perform the real polynomial of indifferent parity resulting from the Chebyshev expansion. }
\label{fig_circuit}
\end{figure*}

The quantum circuit used to perform VCC for a $q$-qubit system is shown in Fig.~\ref{fig_circuit}. The parity of the polynomials generated by a QSVT circuit is restricted; a single QSVT function can return only even or odd polynomials.  Therefore, it is necessary to combine two QSVT circuits, each responsible for even or odd polynomials, to realize the Chebyshev expansion. This combination was achieved using a linear combination of unitaries (LCU)~\cite{M2012}, requiring two ancilla qubits. Additionally, additional $m$-ancilla qubits are required to realize block-encoding \cite{Low2016a,Camps2022,Camps2022a} of $\frac{1}{2\kappa} ( \T + \Td)$ in QSVT.
 After the Chebyshev expansion of $e^{\frac{1}{2} (\T + \Td)}$, the disentangled UCC part $\prod_{\mu} e^{\frac{1}{2}(T_{\mu} - \Td_{\mu})}$ may be applied to the system state, finalizing the HC$^d$-VCC ansatz.

\section{Computational Details}

Here, we provide the computational details of the proof-of-principle numerical simulations of our methods. We implemented the circuit shown in Fig.~\ref{fig_circuit} in Python using Pennylane, version 0.34.0~\cite{Bergholm2018}.  
The angles for the QSVT were determined using the algorithm implemented in Pyqsp, version 0.1.6~\cite{Martyn2021,chao2020finding, Haah2019product, Gilyen2018}. 
It should be noted that, in this manuscript, the exact $e^{\frac{1}{2}{(\T - \Td)}}$ matrix was used instead of the disentangled UCC ansatz.

We computed the potential energy curves (PECs) of $\rm{N}_2$, linear $\rm{H}_4$ and $\rm{H}_6$ molecules to assess the accuracy of our methods. All calculations utilized the minimal basis set STO-3G, with the HF state serving as the reference state. Variational parameters were optimized using the L-BFGS-B algorithm~\cite{L_BFGS_B}. For comparison, we performed full-configuration interaction (FCI) or exact diagonalization. When using the active space approximation, the Complete Active Space Configuration Interaction (CASCI) was employed to obtain reference energies. Classical algorithms were executed using the PySCF program, version 2.3.0~\cite{Sun2020}.

\section{Results and Discussion} \label{seq_results}

%%%%%%%%%%%%%%%%%%%%%%%%%%%%%%%%%%%%%%%%%%%%%%%%%%%%%%%%%%%%%%%%%%%%%%%%%%%%%%%%

\subsection{Chebyshev approximation of VCCSD} \label{sec_cheby_VCCSD}

\begin{figure*}[!tb]
\includegraphics[width=\textwidth]{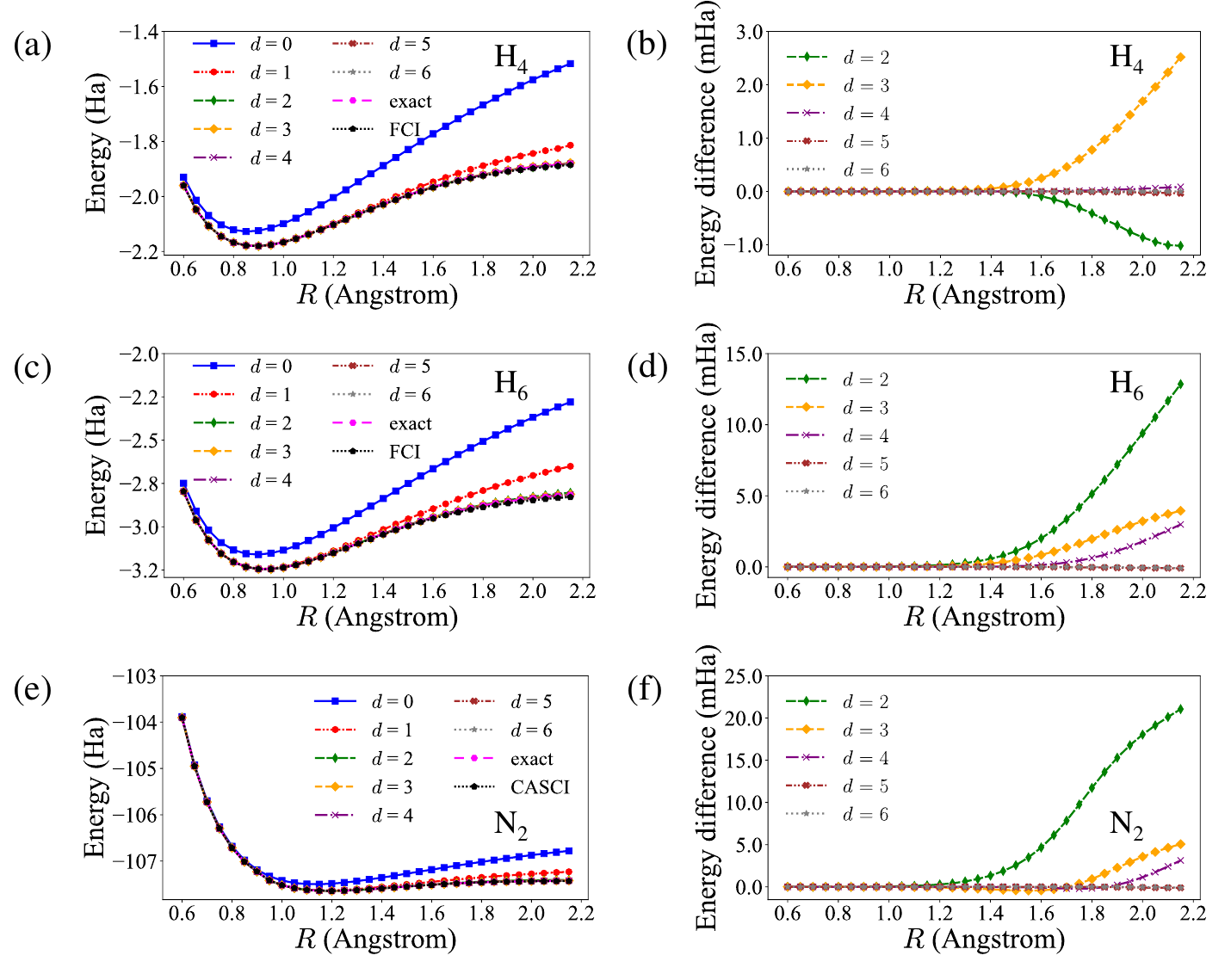}
 \caption{Convergence of C$^d$-VCCSD to exact VCCSD as the degree of truncation in the Chebyshev expansion. The three rows correspond to the results for linear H$_4$, H$_6$, and N$_2$, respectively. The left panels display the PECs of C$^d$-VCCSD, exact VCCSD, and FCI/CASCI for each molecule. The right panels illustrate the energy difference between C$^d$-CCSD and exact VCCSD. The horizontal axes represent the inter-atomic distance $R$ between neighboring atoms. All calculations use the STO-3G basis set, with the active space for N$_2$ set to (6e, 6o).}
 \label{fig_chebyshev}
\end{figure*}

First, we discuss the accuracy of the truncated Chebyshev expansion VCCSD, denoted as C$^{d}$-VCCSD, compared to the exact VCCSD and its dependence on the degree $d$. The simulation results are presented in Fig.~\ref{fig_chebyshev}. The C$^{0}$-VCCSD and C$^{1}$-VCCSD are equivalent to HF and CISD, respectively. As expected, these two methods show significant deviations from the exact VCCSD. However, the C$^{d}$-VCCSD rapidly converges to the exact VCCSD as $d$ increases. 

The right panels of Fig.~\ref{fig_chebyshev} illustrate the truncation errors for cases where $d \geq 2$. The truncation error increases with the interatomic distance $R$ but decreases rapidly with the increase of the degree of truncation.
For the linear H$_4$ molecule, the C$^{4}$-VCCSD shows slight errors. Since H$_4$ is a 4-electron system, the C$^{4}$-VCCSD is expected to be exact. These observed errors likely stem from numerical inaccuracies in the parameter gradients during the optimization process. For H$_6$ and N$_2$, both 6-electron systems, the C$^5$-VCCSD exhibits sub-milliHartree (mHa) errors.  These results suggest that the C$^d$-VCCSD with $d<N$ can sufficiently reproduce the exact VCCSD.

%%%%%%%%%%%%%%%%%%%%%%%%%%%%%%%%%%%%%%%%%%%%%%%%%%%%%%%%%%%%%%%%%%%%%%%%%%%%%%%%%%5

\subsection{Robustness of {\it trotterized} VCCSD} \label{sec_trotter_error}
\begin{figure*}
 \includegraphics[width=\linewidth]{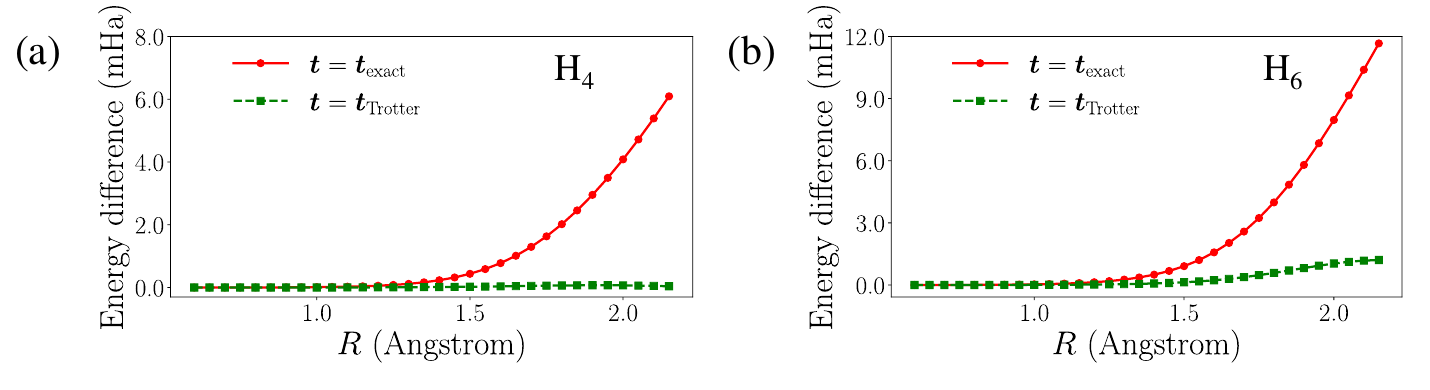}
 \caption{Energy errors in \textit{trotterized} VCCSD using amplitudes optimized for exact VCCSD, ${\bm t} = {\bm t}_{\rm exact}$, and \textit{trotterized} VCCSD, ${\bm t} = {\bm t}_{\rm Trotter}$. Panels (a) and (b) represent the results for linear H$_4$ and H$_6$, respectively. The distance $R$ specifies the interatomic distance between neighboring hydrogens. All calculations were performed using the STO-3G basis set.}
 \label{fig_Trotter_error}
\end{figure*}

Next, we validate the {\it trotterized} VCCSD ansatz in Eq.~(\ref{eq_one_trotter_step}). Fig.~\ref{fig_Trotter_error} compares the {\it trotterized} VCCSD PECs with the exact VCCSD. Note that the {\it trotterized} VCCSD does not employ a Chebyshev expansion, so any deviation from the exact VCCSD is solely due to trotterization. When using the parameters from the exact VCCSD ansatz, the deviation increased in the strongly correlated regime.
However, the variationally optimized {\it trotterized} VCCSD significantly reduced the error, indicating that most Trotter errors can be absorbed into the variational parameters. 
A similar observation has been made in the context of the UCC ansatz \cite{Barkoutsos2018}.
Nonetheless, for longer bond lengths, the Trotter error remains non-negligible. This error may be further reduced using a higher-order Trotter-Suzuki approximation~\cite{Hatano2005,Berry2007}, though at the expense of increased computational costs.

\subsection{Numerical verification of HC$^{d}$-VCCSD} \label{subsection_VCC}

\begin{figure*}
 \includegraphics[width=\linewidth]{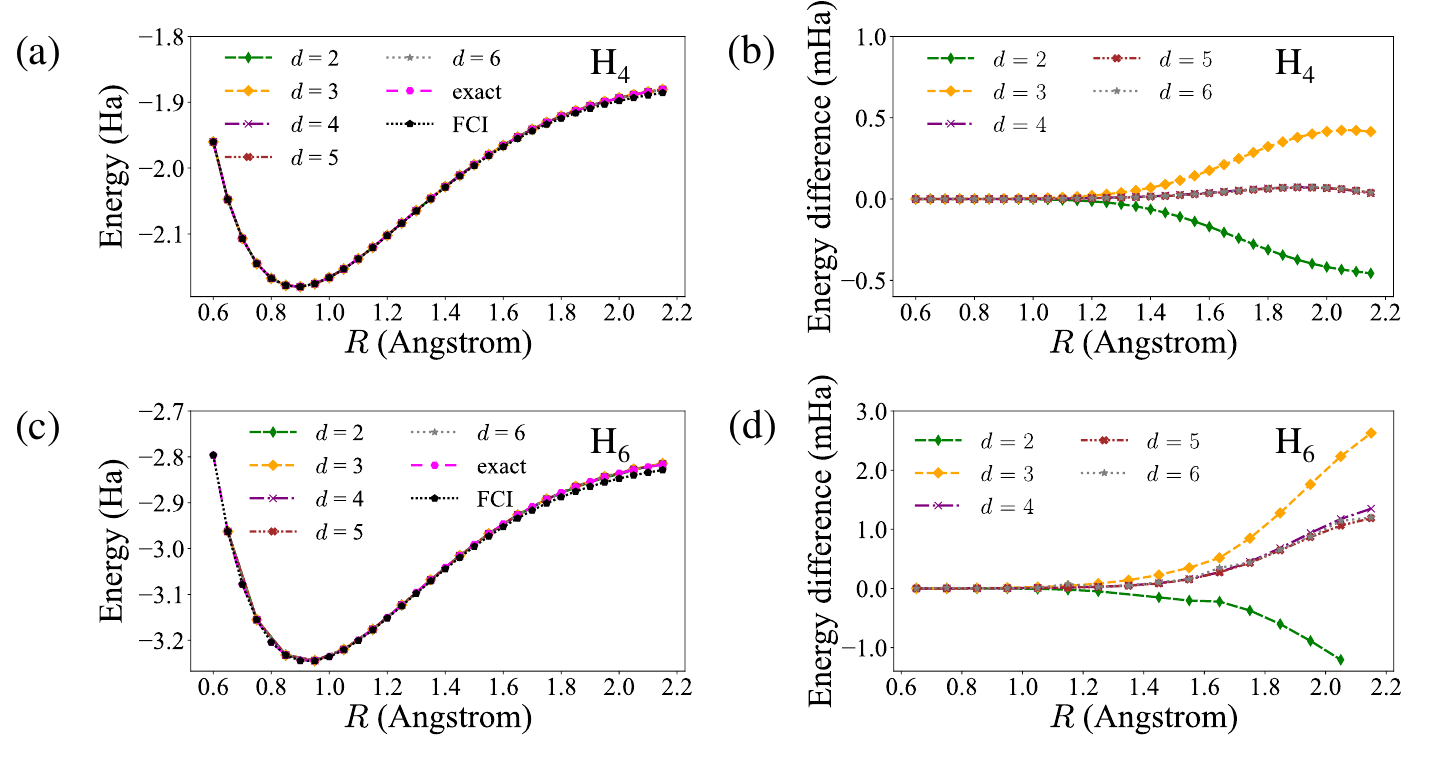}
 \caption{Convergence of HC$^d$-VCCSD to exact VCCSD as the degree of truncation in the Chebyshev expansion increases from $d=2$ to $d=6$. HC$^d$-VCCSD was implemented using QSVT.
 The two rows correspond to the results for linear H$_4$ and H$_6$.
The left panels display the PECs of HC$^d$-VCCSD, exact VCCSD, and FCI for each molecule. The right panels illustrate the energy difference between HC$^d$-CCSD and exact VCCSD. The distance $R$ specifies the interatomic distance between neighboring hydrogens. All calculations were performed using the STO-3G basis set.}
 \label{fig_qsvt}
\end{figure*}

Fig.~\ref{fig_qsvt} shows the performance of the HC$^d$-VCCSD ansatz.
The PECs of the linear H$_4$ and H$_6$ molecules have been simulated. 
The errors in the HC$^d$-VCCSD ansatz stem from two sources: the Chebyshev expansion errors (Fig.~\ref{fig_chebyshev}) and the Trotterization errors (Fig.~\ref{fig_Trotter_error}). For the H$_4$ molecule, where Trotterization errors are negligible, the Chebyshev expansion error is smaller in HC$^d$-VCCSD than in C$^d$-VCCSD. For instance, at an H-H distance of 2.0 Angstrom, the HC$^2$-VCCSD error is approximately 1/4 of the C$^2$-VCCSD error. This result is because HC$^d$-VCCSD does not use the Chebyshev expansion for the disentangled UCC ansatz. In the H$_6$ molecule, for $d=4$ or higher, the main source of error is Trotterization, with the Chebyshev expansion error being small and non-dominant.

\section{Conclusions} \label{seq_conclustion}

In this study, we proposed an approach to implementing the variational coupled cluster (VCC) theory on quantum computers by expanding the exponential cluster operator using Chebyshev polynomials. We introduced the Chebyshev approximated VCC (C$^d$-VCC) and investigated its accuracy and convergence with respect to the truncation degree $d$. Numerical simulations demonstrated that the C$^d$-VCC rapidly converges to the exact VCC with increasing $d$, and the C$^d$-VCC with $d < N$, where $N$ is the number of electrons, reproduced the exact VCC with sufficient accuracy.

Implementing the C$^d$-VCC itself on quantum computers is not straightforward because the cluster operator is not diagonalizable. When an operator is not diagonalizable, quantum
algorithms such as quantum singular value transformation (QSVT) cannot be directly applied. 
To address this issue, we developed the Hermitian-part Chebyshev approximated VCC (HC$^d$-VCC) ansatz, which decomposes the cluster operator into anti-Hermitian and Hermitian parts. 
The anti-Hermitian part is represented by the well-known disentangled unitary coupled cluster (UCC) ansatz, 
while the Hermitian part is approximated using the Chebyshev expansion. 
For Hermitian operators, their Chebyshev polynomials can be realized on quantum circuits using the QSVT technique. Numerical simulations revealed that the HC$^d$-VCC ansatz can effectively reduce the Chebyshev expansion error compared to the C$^d$-VCC, as it does not employ the Chebyshev expansion for the UCC part.

Implementing non-unitary CC wave functions on quantum computers has the potential to simplify the process of initial state preparation on quantum devices, as it is expected that the circuit parameters with small rotation angles can be determined on classical computers~\cite{Hakkaku2022}. Furthermore, this method may also facilitate the efficient extraction of quantum state information, which is essential for post-processing tasks on classical computers following quantum computations. For example, the quantum circuit parameters (i.e., coupled cluster amplitudes) in the active space can be directly used in the framework of tailored or externally-corrected CC methods with quantum inputs~\cite{Scheurer2024, Erhart2024}, efficiently considering the electron correlations ignored in quantum computing on a classical computer. Existing such post-processing methods require many repeated measurements, while the present method allows us to extract the cluster amplitudes without any measurement.

\begin{acknowledgments}
This project was supported by funding from the MEXT Quantum Leap Flagship Program (MEXTQLEAP) through Grant No. JPMXS0120319794, and the JST COI-NEXT Program through Grant No. JPMJPF2014. The completion of this research was partially facilitated by the JSPS Grants-in-Aid for Scientific Research (KAKENHI), specifically Grant No. JP23H03819.
\end{acknowledgments}
\appendix

\bibliography{main.bib}% Include main.bib file
\end{document}